\documentclass[prl,floatfix,twocolumn,showpacs]{revtex4}
\usepackage{epsfig,psfrag,amsmath,amssymb,float}
\usepackage{color}
\begin{document}
\newcommand{\bc}{\begin{center}}
\newcommand{\ec}{\end{center}}
\newcommand{\be}{\begin{equation}}
\newcommand{\ee}{\end{equation}}
\newcommand{\beqn}{\begin{eqnarray}}
\newcommand{\eeqn}{\end{eqnarray}}

\title{Critical properties of doped coupled spin-Peierls chains}

\author{Nicolas Laflorencie$^{(1,2)}$}
\author{Didier Poilblanc$^1$}
\author{Manfred Sigrist$^3$}
\affiliation{$^{(1)}$Laboratoire de Physique Th\'eorique, Universit\'e
Paul Sabatier, F-31062 Toulouse, France}
\affiliation{$^{(2)}$Department of Physics \& Astronomy, University of British Columbia, Vancouver, B.C., V6T 1Z1 Canada}
\affiliation{$^{(3)}$Institute of Theoretical Physics,
ETH H\"onggerberg, CH-8093 Z\"urich, Switzerland}
\date{\today}
\begin{abstract}

Using numerical Real Space Renormalisation Group methods as well
as Stochastic Series Expansions Quantum Monte Carlo simulations
a generic model of diluted spin-$\frac{1}{2}$ impurities
interacting at long distances is investigated.
Such a model gives a generic description of coupled dimerized
spin-Peierls chains doped
with non-magnetic impurities at temperatures lower than the spin gap.
A scaling regime with temperature power-law behaviors in
several quantities like the uniform or staggered susceptibilities
is identified and interpreted in terms of large clusters of correlated spins.
\end{abstract}

\maketitle

%============ BODY OF PAPER ===================================

Low dimensional gapped quantum magnets have
attracted a lot of interest in condensed matter physics for many
years. The possibility of doping such systems has lead to an
extremely rich emerging field.
The discovery of the first non-organic
spin-Peierls compound CuGeO$_3$ \cite{Hase93} and, soon later,
its doping with static non-magnetic impurities
realized by direct substitution of a small fraction of
copper atoms by zinc \cite{Hasedop93} or magnesium
\cite{dopedCuGe03} atoms offered a new challenge for
the theorist and an ideal experimental system to
test new theoretical concepts.

Our aim here is to analyse the low temperature properties of a
typical two-dimensional array of coupled, frustrated and dimerized
antiferromagnetic (AF) spin-$\frac{1}{2}$ chains doped with
non-magnetic impurities. For that purpose we use the
low-temperature effective model derived in previous work: each
non-magnetic dopant releases a spin-$\frac{1}{2}$, localized in
its vicinity~\cite{Sorensen}. These effective spins become the
only remaining low-energy degrees of freedom at temperatures lower
than the spin gap i.e. the energy scale of condensation of the
background spins into nearest-neighbor dimers. The corresponding
effective Hamiltonian describes interacting spins $\frac{1}{2}$
randomly distributed on a square lattice (of size $L\times L$),
\be {\mathcal{H^{\rm{eff}}}}=\sum_{{\bf r}_1,{\bf r}_2}
\epsilon_{{\bf r}_1} \epsilon_{{\bf r}_2}J({\bf r}_1-{\bf r}_2)
{\bf S}_{{\bf r}_1}\cdot {\bf S}_{{\bf r}_2}, \label{EffHam} \ee
where the occupation number $\epsilon_{\bf r}$ takes random values
$1$ ($0$) with probability $x$ ($1-x$), $x=N_s/L^2$ is the dopant
concentration, $N_s$ the number of spins $\frac{1}{2}$. The
effective interaction $J({\bf r})$ computed by Lanczos exact
diagonalizations from the original microscopic model~\cite{prl03}
bears important properties: (i) it has opposite signs on the two
(relative) sublattices and, hence, is {\it{non-frustrating}} in
nature and (ii) it displays a typical (spatially anisotropic)
exponential (long distance) behavior, characterized essentially by
two length scales, $\xi_{\parallel}$ and $\xi_{\perp}$,
\be
J(\Delta x ,\Delta y)\propto -(-1)^{\Delta x +\Delta y}
\exp(-\tilde\Delta x/\xi_{\parallel}-\Delta y/\xi_{\perp}),
\label{eq:Jeff}
\ee
where $\tilde\Delta x =\Delta x+x_{\rm max}$ (see later)
and $\Delta x =|x_1 -x_2|$ ($\Delta y =|y_1 -y_2|$) is the
separation between 2 dopants in the longitudinal (transverse)
direction. Typically, we assume hereafter $\xi_{\parallel}=2.5$ and $\xi_{\perp}=1$.
Note that Eq.(\ref{eq:Jeff}) is generic for most doped
spin gapped systems~\cite{prl03,Sigrist96,Wessel01}.
%Although the exact
%form of $J(\Delta x ,\Delta y)$ at short distances is not crucial
%at low $x$,
In order to use a consistent and realistic description
of the actual experimental compound, we
shall use also the specific form
derived in Ref.~\cite{prl03} at short distances for an AF
interaction i.e. when
$\Delta x+\Delta y$ is odd~\cite{note0}:
$J(\Delta x,\Delta y)\propto \Delta x$ up to a maximum at
distance $x_{\rm max}=2\xi_\parallel$ ($x_{\rm max}=0$
for $\Delta x+\Delta y$ even).
At larger distances, the asymptotic form of Eq.(\ref{eq:Jeff})
is used.

As supported by experiments~\cite{Grenier98}, the present framework implies
naturally that the low-temperature uniform susceptibility scales
with the impurity concentration and displays a Curie-like behavior
$\chi=C/T$. As argued in \cite{Sigrist96}, due to the presence of
AF as well as ferromagnetic (F) couplings~\cite{Sigrist1}, large
(weakly coupled) clusters of spins form. Using a simple classical
random walk argument, the effective spin $S^{\rm{eff}}$ of the
clusters containing $n$ spins is simply given, on average, by
$\langle S^{\rm{eff}}\rangle = \frac{1}{2} \langle n \rangle
^{1/2}$. Consequently, the Curie constant (per spin) should
saturate to a value equal to $\frac{1}{3}(\langle
S^{\rm{eff}}\rangle )^2 / \langle n \rangle=1/12$ when $T\to 0$.
Preliminary numerical results are in agreement with this
prediction~\cite{LPS03}. Here we shall use both numerical Real
Space Renormalization Group (RSRG) \cite{BL82,Fisher94} supplemented by Stochastic
Series Expansion (SSE) Quantum Monte Carlo (QMC)~\cite{Sandvik1}
to investigate the critical properties as one approaches the zero
temperature limit.

For a single spin-$\frac{1}{2}$ chain with random AF couplings,
Fisher demonstrated the existence of a universal
fixed point for the
renormalization group transformation~\cite{Fisher94}, thus providing a strong
argument in favor of such a procedure. In the case of both
random F and AF couplings,
while there is no exact solution, the existence of a
fixed point was shown unambiguously by Westerberg et
al.~\cite{Sigrist1} using a RSRG approach. Its validity was
discussed very carefully in \cite{Sigrist1} and the predictions
were successfully checked by QMC~\cite{MonteCarloFAF}.
In the similar 2D case we
address here, the distribution of couplings $P(|J|)$
behaves like $1/|J|$ and does not depend on the dilution $x$
(apart from logarithmic factors and a finite size
cutoff $J_{\rm{min}}$).
The RSRG is expected to work particularly well for such a
singular distribution.
%%%%%%%%%%%FIG:%%%%%%%%%%%%%%%%%
\begin{figure}[!ht]
\bc
\epsfig{file=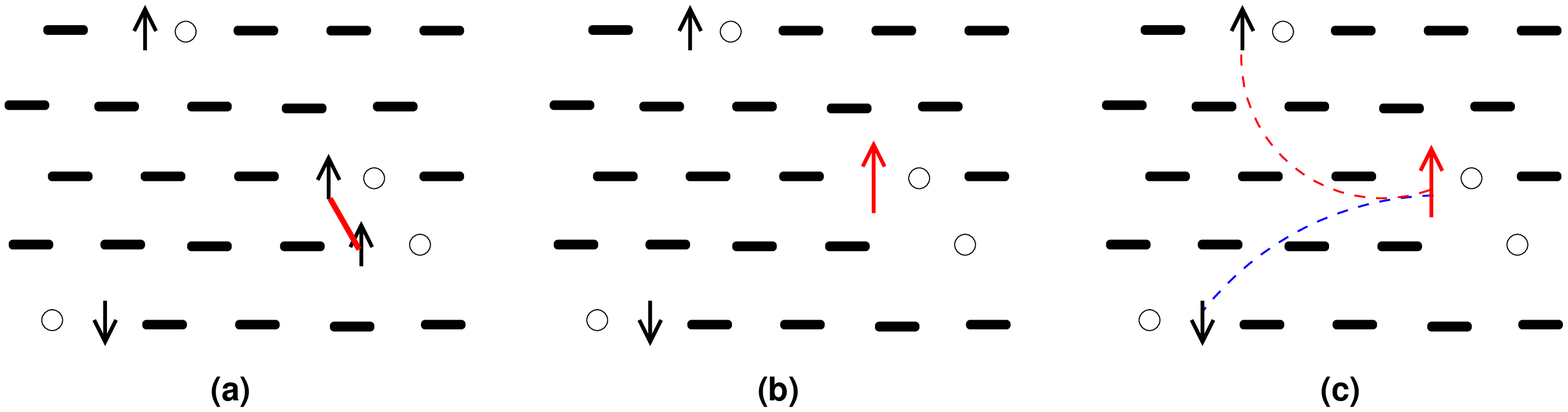,width=8.4cm,height=2.5cm,clip}
\caption{Schematic picture of the doped spin-Peierls system.
Thick bonds stand for dimers and non-magnetic impurities
(released spins $\frac{1}{2}$) are represented by open circles
(black arrows). The initial RSRG step is illustrated starting from
a typical configuration with 4 impurities:
(a) The strongest coupled pair is identified (red line).
(b) This pair, e.g. ferromagnetic here, is
replaced by a spin $S=1$ (red arrow). (c) The
couplings with all other spins (dashed lines) are renormalized.}
\label{fig:CurieRSRG} \ec
\end{figure}
%%%%%%%%%%%%%%%%%%%%%%%%%%%%%%%%%%%

Following the pioneering work of Bhatt and Lee~\cite{BL82},
we extend the RSRG scheme to hamiltonian (\ref{EffHam})
with F and AF long-distance couplings~\cite{Sigrist1,Melin}. Let
us define the effective interaction
as $J_{i,j}$ where $i$ and $j$ label the {\it randomly distributed
spins} and run from $1$ to $N_s$. One single RG step is described
as follows : 1) Identify the most strongly coupled pair of spins
($S_1,S_2$) i.e. with the largest energy gap $\Delta_{1,2}$, $
\Delta_{1,2}=J_{1,2}(1+\left|S_1 -S_2 \right|)
~{\rm{if}}~J_{1,2}>0~{\rm{(AF)}}$ and $\Delta_{1,2}=-J_{1,2}(S_1
+S_2) ~{\rm{if}}~J_{1,2}<0~{\rm{(F)}}$. Note that $\Delta_{1,2}$
defines the energy scale of the transformation.
%(which will, in fact, play the role of the temperature. 
2) Replace it by an
effective spin $S^{'}=\left|S_1 -S_2 \right|$ if the coupling is
AF or $S^{'}=S_1 +S_2$ in the F case. 3) Renormalize all the
magnetic couplings with the following rules : (i) If $S^{'}\neq
0$, as given by a first order perturbation theory, the new
couplings between $S^{'}$ and all the other spins ($S_3,
S_4,...,S_{N_s}$) are set to
$$
{\tilde{J}}_{(S^{'},S_{i})}=J_{1,i}~c(S_1,S_2,S^{'})+J_{2,i}~c(S_2,S_1,S^{'}),
\hfill {\rm with}$$
$$
c(S_1,S_2,S^{'})=\frac{S^{'}(S^{'}+1)+S_1(S_1+1)-S_2(S_2+1)}{2S^{'}(S^{'}+1)}.
$$
(ii) If $S^{'}=0$, the pair ($S_1,S_2$) is frozen.
Using a cluster approximation~\cite{BL82}
that involves only the extra pair ($S_3,S_4$)
 the most strongly coupled to
$S_1$ and $S_2$ and a second order perturbation,
the coupling $J_{3,4}$ is renormalized as
$$
{\tilde{J}}_{3,4}=J_{3,4}+\frac{2S_1(S_1+1)}{3J_{1,2}}(J_{1,3}-J_{2,3})(J_{2,4}-J_{1,4}).
$$
The same procedure is then reiterated. We also check that the RSRG preserves the non-frustrated character of the problem.

\begin{figure}[!ht]
\bc 
\psfrag{S}{$\langle S^{\rm{eff}} \rangle$} 
\psfrag{D}{$\langle \Delta_0 \rangle$} 
\psfrag{n}{\tiny{$\langle n \rangle$}} 
\psfrag{d}{\tiny{$\langle \Delta_0 \rangle$}}
\epsfig{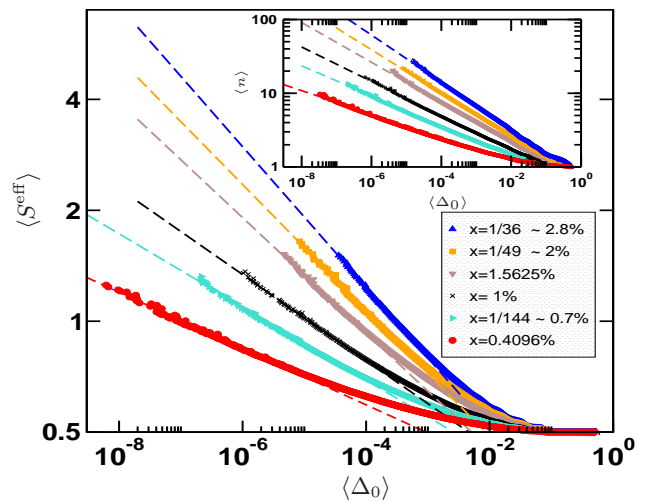}
\caption{Average effective spin $\langle S^{\rm{eff}}\rangle$ of the
clusters of active spins vs the energy scale $\langle \Delta_0
\rangle$ for six different concentrations $x$ indicated on the
plot. Numerical RSRG data obtained for $N_s =1024$ spins over more
than $10^4$ samples. Inset: for the same samples, average number
$\langle n \rangle$ of initial spins-$\frac{1}{2}$ per cluster vs
$\langle\Delta_0 \rangle$. Dashed lines are power-law fits (see
text).} \label{fig:SpinEffRSRG} \ec
\end{figure}

\begin{figure}[!ht]
\bc 
\psfrag{A}{{\Large{$x$}}}
\psfrag{cr}{{\bf{Critical exponents}}}
\epsfig{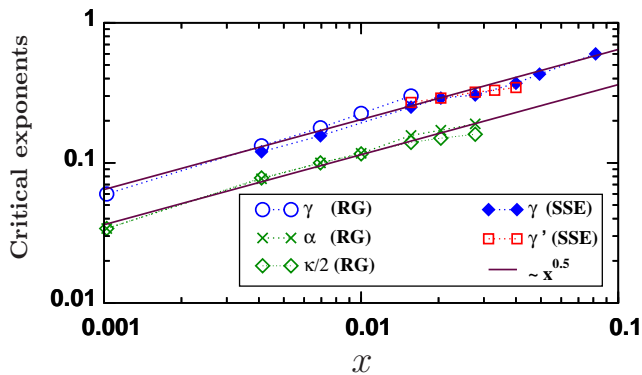}
\caption{Exponents
$\alpha(x)$, $\gamma(x)$ and $\gamma'(x)$ extracted from various SSE and RSRG data as indicated on the plot (see text for details). Straight lines are $\sim \sqrt{x}$.}
\label{fig:alphax} 
\ec
\end{figure}

Due to the presence of both F and AF couplings, clusters with
large effective spins are created during the procedure similarly
to what occurs in the 1D random F-AF spin-$\frac{1}{2}$
chain~\cite{Sigrist1}. At each RG step, the energy scale
$\Delta_0$ decreases and both the number of inactive spins frozen
into singlets and the number of clusters build from a large number
$n$ correlated spins-$\frac{1}{2}$, increase. The aforementioned random
walk argument predicts that, the average number $\langle n
\rangle$ of spins-$\frac{1}{2}$ inside clusters and their average spin
magnitude $\langle S^{\rm{eff}}\rangle $ should be related by
$\langle S^{\rm{eff}}\rangle \sim \langle n \rangle ^{1/2}$ at low
enough temperatures. Therefore we expect the effective spin of
these clusters to grow monotonously as the energy scales down. We
have analyzed this process using the RSRG scheme to compute both
$\langle S^{\rm{eff}}\rangle $ and $\langle n \rangle$ as a
function of $\langle \Delta_0\rangle$. This is shown in
Fig.~\ref{fig:SpinEffRSRG} which clearly demonstrates the
formation of large moments. Moreover, power-law divergences like
$\langle S^{\rm{eff}}\rangle \sim \langle\Delta_0 \rangle
^{-\alpha (x)}$ and $\langle n \rangle \sim \langle\Delta_0
\rangle ^{-\kappa (x)}$ are observed with $\kappa \simeq 2
\alpha$. We have plotted the behavior of these exponents vs $x$ in
Fig.~\ref{fig:alphax}. Interestingly enough, we find that $\alpha$
depends on $x$ in contrast to the random F-AF spin chain for which
$\alpha=0.22\pm 0.01$~\cite{Sigrist1}. Note however that the
universality class identified in \cite{Sigrist1} as a RG fixed
point occurs only for initial gap distributions less singular than
$P_c (\Delta)\sim \Delta ^{-y_c}$, with $0.65\lesssim y_c \lesssim
0.75$. For more singular distributions, as it is the case here,
critical exponents are not universal anymore, at least in
1D~\cite{Sigrist1}.
%
%%%%%%%%%%%FIG:SPINEFFECT.RSRG%%%%%%%%%%%%%%%%%
%%%%%%%%%%%%%%%%%%%%%%%%%%%%%%%%%%%

We now turn to a brief discussion about the validity of the RSRG
approach w.r.t. the large (formally infinite) connectivity of the
model. Whereas the moments eventually order AF at zero temperature
for arbitrary small concentration~\cite{LPS03}, at finite
temperatures the physics is dominated by short range couplings,
i.e. inside a domain defined by $\Delta x \le \xi_{\parallel}$ and
$\Delta y\le \xi_{\perp}$. Consequently, one could define a
"short-range connectivity number" $z_{sr}$ which is, for a given
spin, the average number of neighboring spins belonging to such a short range couplings region. It
is easy to check that
$z_{sr}={\mathcal{N}}_{sr}x(1-N_{s}^{-1})\simeq
{\mathcal{N}}_{sr}x$ where ${\mathcal{N}}_{sr}$ is the number of
sites contained in the short-range region. For the present model
${\mathcal{N}}_{sr}\simeq 65$ sites as seen numerically.
The evolution under the RSRG scheme depends on the value of
$z_{sr}$, hence on $x$. When, let us say, $z_{sr}<1$ (i.e.
$x<1.55\%$), the probability to have extra spins strongly coupled
to the most strongly coupled pair ($S_1,S_2$) remains very small,
so that the necessary condition $J_{1,2}\gg J_{1,i}, J_{2,i}$,
$\forall i>2$ is fulfilled. On the other hand, if $z_{sr}>2$ (i.e.
$x>3.1\%$) a ``percolation threshold'' is reached and one expects
the formation of strongly coupled clusters reaching the system
size.
%
%%%%%%%%%%%%%%%%%%%%%%%%%%%%%%%%%%%%%%%%%%%%%%%%%%%%%%%%%%%%
\begin{figure}[!ht]
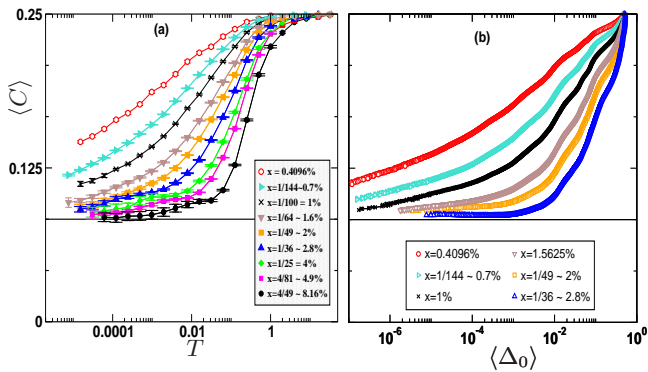

\bc
\begin{minipage}{0.495\columnwidth}
\vskip -0.14cm
\psfrag{C}{$\langle C \rangle$}
\psfrag{T}{$T$}
\epsfig{file=CurieSSE.eps,width=4.4cm,height=4.65cm,clip}
\end{minipage}
\begin{minipage}{0.49\columnwidth}
\psfrag{D}{$\langle \Delta_0 \rangle$}
\epsfig{file=CurieRSRG.eps,width=4cm,height=4.73cm,clip}
\end{minipage}
\caption{Curie constant per spin $\langle C \rangle$ plotted vs the energy scale. (a) Quantum Monte Carlo SSE
results shown vs $T$ for $N_s =256$ spins and 9 different
concentrations $x$ indicated on the plot. Disorder is performed on
$10^3$ to $10^4$ samples. (b) Numerical RSRG results shown for
$N_s =1024$ spins and 6 different concentrations $x$ as indicated
on the plot, vs the RG energy scale $\langle\Delta_0 \rangle$. Error
bars are typically smaller than symbol sizes, the number of
samples always exceeding $10^4$.  The full line correspond to the
saturation value of $1/12$.}
\label{Curie} 
\ec
\end{figure}
%%%%%%%%%%%%%%%%%%%%%%%%%%%%%%%%%
%

\begin{figure}[!ht]
\bc 
\psfrag{C}{$\langle C \rangle-\frac{1}{12}$}
\psfrag{A}{\tiny{(a) $x\simeq 0.41\%$}}
\psfrag{B}{\tiny{(b) $x\simeq 0.7\%$}}
\psfrag{D}{\tiny{(c) $x=1\%$}}
\psfrag{E}{\tiny{(d) $x\simeq 1.56\%$}}
\psfrag{F}{\tiny{(e) $x\simeq 2\%$}}
\psfrag{G}{\tiny{(f) $x\simeq 2.8\%$}}
\psfrag{T}{$T,~\langle \Delta_0 \rangle$}
\epsfig{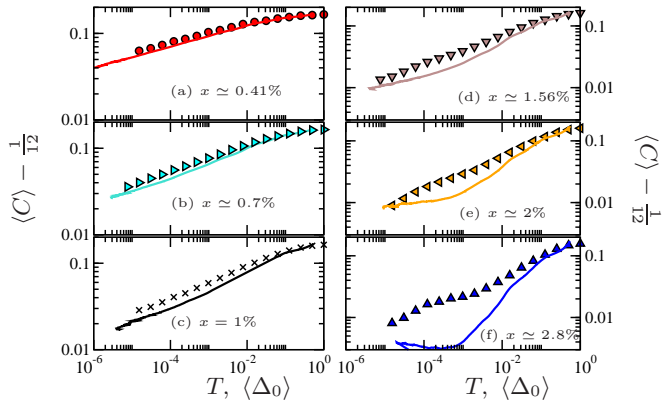}
\caption{Direct comparisons
between RSRG (full lines) and SSE simulations (symbols) of $\langle C
\rangle-1/12$ are shown for the 6 different concentrations indicated by (a), (b),..., (f) vs the RG energy scale $\langle\Delta_0 \rangle$ or the SSE temperature $T$.}
\label{Compare} \ec
\end{figure}

Since the above arguments are still qualitative \cite{NoteRG}, we have
confronted the RSRG results  to QMC data obtained on the same
model and for the same parameters. The SSE method \cite{Sandvik1}
supplemented by the $\beta$-doubling scheme~\cite{Sand2}, already
used in~\cite{LPS03}, can reach extremely low temperatures.
Figs.~\ref{Curie} show results for the Curie constant per spin
obtained with both methods. The RSRG computation of $C$ is
performed, at each RG step, using the formula
$C=\frac{1}{3N_s}\sum_{\sigma}N_{\sigma}\sigma (\sigma +1)$ where
$N_{\sigma}$ is the number of active effective spins of size
$\sigma$, the data being then averaged over disorder. We have also
checked that finite size effects are negligible when $N_s \ge 256$
and we have chosen $N_s =1024$ in most computations. We observe a
qualitative agreement between Fig.\ref{Curie}(a) and
Fig.\ref{Curie}(b) where, at high temperatures, the spins behave
as paramagnetic free magnetic moments (giving a Curie constant of
$\frac{1}{4}$ per spin) and where saturation to $\frac{1}{12}$ is
observed at low $T$, the spins being correlated inside large
clusters. At small concentration, the agreement becomes even
quantitative, as seen in the Fig.\ref{Compare} where a comparison between RSRG and SSE is shown for the six lowests values of $x$.
%%%%%%%%%%%FIG:COMPAR.SSE-RSRG.CURIE%%%%%%%%%%%%%%%%%
%%%%%%%%%%%%%%%%%%%%%%%%%%%%%%%%%%%%%%%%%%%%%%%
The agreement remains quite good for
larger concentration up to $x\simeq 1.56\%$ corresponding to
$z_{sr}\simeq 1$, hence corroborating the qualitative argument
stated above.

%Figs.\ref{Curie} show that $\langle C \rangle$ decreases faster
%for larger $x$. In addition, one observes in Fig.\ref{Curie} (B)
%for the largest concentrations $x\ge 1/36$ an intermediate regime
%of $T$ with a small {\it{plateau}} in the Curie constant, after a
%rapid fall-off. On the other hand, for smaller concentrations $x
%\le 1/49$, the plateau regime is not observed and $\langle C
%\rangle$ follows a smooth behavior. This plateau regime can be understood from
%simple qualitative arguments invoking two regimes; (i) A transient
%regime where clusters of correlated spins
%(within a "correlated region" of typical size $\sim \xi_\parallel \xi_\perp$)
%rapidly form hence reducing strongly the
%susceptibility. This occur at lower $T$ for smaller $x$ since the
%``typical'' coupling between nearby impurities
%decreases rapidly with the characteristic
%length scale set by their mean separation; (ii) A
%regime where the clusters are well formed and behave
%eventually like effective spins once, at sufficiently low
%temperature, their internal degrees of freedom are frozen. Such
%magnetic moments might start to interact only at energy scales
%lower than the magnitude of the ``typical'' coupling, hence
%leading to a plateau on a limited temperature range as seen for
%the largest concentrations studied. This can also be interpreted
%as an "effective energy gap" in the distribution of effective
%couplings.
We now turn to the analysis of the scaling regime. At very low
temperatures, the quantum corrections $\langle C(T)\rangle -1/12$
are expected to behave like $T^{\gamma}$ where $\gamma=\alpha$ in 1D~\cite{MonteCarloFAF}.
%%%%%%%%%%%FIG:CURIE.COLLAPSE.SSE%%%%%%%%%%%%%%%%%
\begin{figure}[!ht]
\bc 
\psfrag{C}{Curie constant}
\psfrag{T}{$T^{\gamma_{\rm{_{SSE}}}}$}
\psfrag{D}{$\langle \Delta_{0}\rangle ^{\gamma_{\rm{_{RG}}}}$}
\epsfig{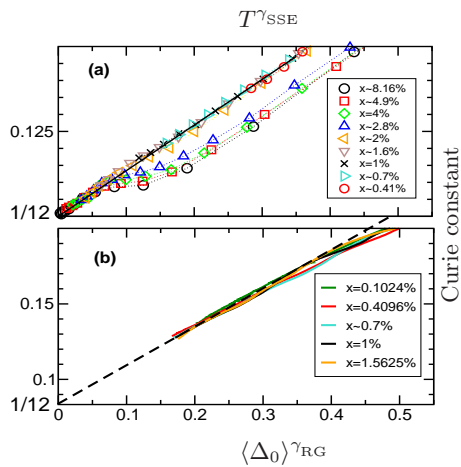}
\caption{Curie constant $\langle C \rangle$ plotted vs
$T^{\gamma_{\rm{_{SSE}}}}$ for the QMC data in (a) and vs $\langle \Delta_{0}\rangle ^{\gamma_{\rm{_{RG}}}}$ for the RSRG data in (b), in order to obtain the best data collapse at low energy.} 
\label{fig:SSEColl} 
\ec
\end{figure}
%%%%%%%%%%%%%%%%%%%%%%%%%%%%%%%%%%%%%%%%%%%%%%%
In the present case, the procedure used in Figs.~\ref{fig:SSEColl} to extract
the doping dependence of $\gamma$ from the QMC and RSRG susceptibility data
is the following: First, one estimates the value of $\gamma$ at
the lowest concentration available, i.e. $x_{\rm{min}}\simeq 0.41\%$ for the SSE and $x_{\rm{min}}=0.1\%$ for the RG,  via a
direct reliable power-law fit, giving $\gamma_{\rm{_{SSE}}}=0.12$ and $\gamma_{\rm{_{RG}}}=0.065$ in thoses cases. Then,
the other estimates for larger $x$ are determined in order to
obtain the best collapse of all the data plotted vs $T^\gamma$ on
a universal low temperature curve, as shown in Figs.~\ref{fig:SSEColl}. 
We note that, for large $x$,
deviations at high temperatures can be attributed to a transient
and plateau regimes also visible in Fig.~\ref{Curie}(a) for $x \ge 1/36$~\footnote{Such a behavior can be understood as an effective ``energy gap'' in the distribution of effective couplings, due a short-range connectivity number larger than 1.}.
%%%%%%%%%%%FIG:CHISTAG.SSE%%%%%%%%%%%%%%%%%
\begin{figure}[!ht]
\bc 
\psfrag{C}{$T\times \chi_{\rm{stag}}$}
\psfrag{T}{$T$}
\epsfig{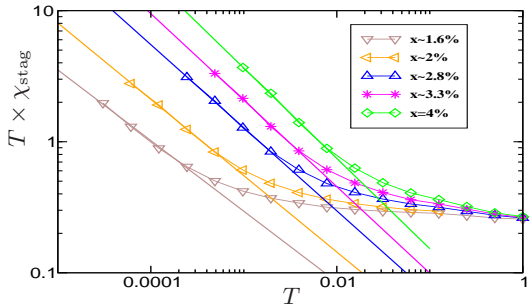}
\caption{$T\langle \chi_{\rm stag}(T)\rangle$ plotted vs
$T$ for five different concentrations. Full
lines are fits corresponding to power-law behaviors $\sim T^{-2\gamma'}$. All data are
computed by QMC using $N_s=256$ random spins and averaged over
disorder.} \label{fig:ChiStag} \ec
\end{figure}
%%%%%%%%%%%%%%%%%%%%%%%%%%%%%%%%%%%%%%%%%%%%%%%
Similarly, one expects for the staggered
susceptibility (per spin), $\langle \chi_{\rm{stag}}(T)\rangle
\propto T^{-1-2\gamma'}$. In 1D, $\gamma'$ is expected to be equal to $\alpha$~\cite{Nagaosa96}, 
but in the present case, by direct fits of the low T (see Fig.~\ref{fig:ChiStag}), we found $\gamma'\simeq\gamma\simeq 2\alpha$.
%
%%%%%%%%%%%%%%%%%%%%
%%%%%%%%%%%%%%
Thoses exponents are plotted in
Fig.~\ref{fig:alphax}. An overall very good agreement is seen
between the different methods, in particular between the estimates of $\gamma$ 
obtained from the analysis of the Curie constant computed by QMC
and RSRG. We stress again that the exponent $\alpha$ deduced from the analysis of
the change of cluster sizes and spins is roughly a factor of
2 smaller than $\gamma$. Interestingly enough,
$\gamma \simeq 2\alpha \propto \sqrt{x}$ in all cases.

To conclude, confronting results
obtained by both state-of-the-art QMC simulations and
numerical RSRG methods,
we have achieved a physical understanding of
the critical properties of a generic 2D model of diluted $S=\frac{1}{2}$ spins
with long-ranged interactions. For small dopant concentration, the RSRG appears to be in excellent agreement with QMC computations.
Power-law temperature behaviors in
several quantities like the uniform or staggered susceptibilities
are revealed and interpreted in terms of large clusters of correlated spins
which also have interesting scaling properties with temperature.
It is remarkable
that such properties are observed above the
$T=0$ ordered magnetic groundstate. Moreover, it is the first example of a two-dimensionnal random magnet exhibiting a large spin phase with a disorder (the concentration $x$) dependence of the critical exponents. Whereas doped CuGeO$_3$ might be a good candidate for the experimental observation of a critical regime, the so large three-dimensionnal ordering temperature prevents such an observation. More strongly diluted samples are necessary to reach the scaling regime.

{\it Acknowledgment:} We are indebted to A.W.~Sandvik for providing
some QMC codes and for valuables comments. DP also thanks T. Giamarchi for usefull discussions.


\begin{thebibliography}{00}
\bibitem{NoteRG} We have also computed the mean gap $\langle \Delta \rangle $ and checked that the ratio $\langle \Delta \rangle / \langle \Delta_0 \rangle $ remains $\le 10^{-2}$.
\end{thebibliography}
\end{document}